%% file: main.tex
\documentclass[a4paper]{llncs}

\usepackage[english]{babel}
\usepackage[utf8x]{inputenc}
\usepackage{multirow}
\usepackage[table]{xcolor}
\usepackage{rotating}
\usepackage{hhline}
\usepackage{wrapfig}
\usepackage{fnpct}
\usepackage[inline]{enumitem}
\usepackage{tikz}
\usetikzlibrary{arrows}
\usetikzlibrary{calc}
\usetikzlibrary{fit}
\usetikzlibrary{decorations.pathreplacing}
\usetikzlibrary{backgrounds}
\usepackage[colorinlistoftodos]{todonotes}
\usepackage{hyperref}
\usepackage{listings}
\usepackage{xcolor}
\usepackage{tcolorbox} 
\usepackage{tabularx}
\usepackage{graphicx}
\usepackage{subcaption}
\usepackage{orcidlink} % For ORCID logos

\definecolor{darkgreen}{rgb}{0.0, 0.5, 0.0}
\definecolor{darkgray}{rgb}{0.1, 0.1, 0.1} % Standard dark color for function names

\title{Developing Retrieval Augmented Generation (RAG) based LLM Systems from PDFs: An Experience Report}

\titlerunning{Developing Retrieval Augmented Generation (RAG) based LLM Systems from PDFs: An Experience Report}

\author{}  % Author section is defined separately below

\institute{}
\begin{document}
\maketitle

% Authors information with ORCID
\begin{center}
    \begin{tabular}{p{5cm} p{5cm}}
        \centering
        \orcidlink{0009-0004-0134-8313} \textbf{Ayman Asad Khan} \\
        Tampere University \\
        \href{mailto:ayman.khan@tuni.fi}{ayman.khan@tuni.fi}
        &
        \centering
        \orcidlink{0009-0005-8907-9898} \textbf{Md Toufique Hasan} \\
        Tampere University \\
        \href{mailto:mdtoufique.hasan@tuni.fi}{mdtoufique.hasan@tuni.fi}
        \\
    \end{tabular}
    \\[1cm] % Space between rows
    \begin{tabular}{p{4cm} p{4cm} p{4cm}}   
        \centering
        \orcidlink{0000-0002-0225-4560} \textbf{Kai Kristian Kemell} \\
        Tampere University \\
        \href{mailto:kai-kristian.kemell@tuni.fi}{kai-kristian.kemell@tuni.fi}
        &
        \centering
        \orcidlink{0000-0002-4401-8013} \textbf{Jussi Rasku} \\
        Tampere University \\
        \href{mailto:jussi.rasku@tuni.fi}{jussi.rasku@tuni.fi}
        &
        \centering
        \orcidlink{0000-0002-4360-2226} \textbf{Pekka Abrahamsson} \\
        Tampere University \\
        \href{mailto:pekka.abrahamsson@tuni.fi}{pekka.abrahamsson@tuni.fi}
    \end{tabular}
\end{center}

\begin{abstract}
This paper presents an experience report on the development of Retrieval Augmented Generation (RAG) systems using PDF documents as the primary data source. The RAG architecture combines generative capabilities of Large Language Models (LLMs) with the precision of information retrieval. This approach has the potential to redefine how we interact with and augment both structured and unstructured knowledge in generative models to enhance transparency, accuracy and contextuality of responses. The paper details the end-to-end pipeline, from data collection, preprocessing, to retrieval indexing and response generation, highlighting technical challenges and practical solutions. We aim to offer insights to researchers and practitioners developing similar systems using two distinct approaches: \textit{ OpenAI's Assistant API with GPT Series} and\textit{ Llama's open-source models}. The practical implications of this research lie in enhancing the reliability of generative AI systems in various sectors where domain specific knowledge and real time information retrieval is important. The Python code used in this work is also available at: \href{https://github.com/GPT-Laboratory/RAG-LLM-Development-Guidebook-from-PDFs}{\textcolor{blue}{GitHub.}}
\end{abstract}


\input{./sections/01_Introduction}
\input{./sections/02_Background}
\input{./sections/03_StudyDesign}
\input{./sections/04_ResultsStepbyStepGuidetoRAG}
\input{./sections/05_Evaluation}
\input{./sections/06_Discussion}
\input{./sections/07_Conclusions}

\newpage

\bibliography{main}
\bibliographystyle{plain}

\newpage
\section*{Appendix}
\label{Appendix}

\begin{enumerate}
    \item Tampere University, “Cost Estimation for RAG Application Using GPT-4o”, Zenodo, Sep. 2024. doi: \href{https://doi.org/10.5281/zenodo.13740032}{10.5281/zenodo.13740032}.
\end{enumerate}

\end{document}

%% file: sections/01_Introduction.tex
\section{Introduction}
\label{section: Introduction}

Large language models (LLMs) excel at generating human like responses, but base AI models can’t keep up with the constantly evolving information within dynamic sectors. They rely on static training data, leading to outdated or incomplete answers. Thus they often lack transparency and accuracy in high stakes decision making. Retrieval Augmented Generation (RAG) presents a powerful solution to this problem. RAG systems pull in information from external data sources, like PDFs, databases, or websites, grounding the generated content in accurate and current data making it ideal for knowledge intensive tasks.

In this report, we document our experience as a step-by-step guide to build RAG systems that integrates PDF documents as the primary knowledge base. We discuss the design choice, development of system, and evaluation of the guide, providing insights into the technical challenges encountered and the practical solutions applied. We detail our experience using both proprietary tools (OpenAI) and open-source alternatives (Llama) with data security, offering guidance on choosing the right strategy. Our insights are designed to help practitioners and researchers optimize RAG models for precision, accuracy and transparency that best suites their use case. 

%% file: sections/02_Background.tex
\section{Background}
\label{section: Background}

This section presents the theoretical background of this study. Traditional generative models, such as GPT, BERT, or T5 are trained on massive datasets but have a fixed internal knowledge cut off based on their training data. They can only generate \textbf{black box} answers based on what they \textbf{know}, and this limitation is notable in fields where information changes rapidly and better explainability and traceability of responses is required, such as healthcare, legal analysis, customer service, or technical support. 

\subsection{What is RAG?}
\label{subsection: What is RAG }

The concept of Retrieval Augmented Generation (RAG) models is built on integrating two core components of NLP: Information Retrieval (IR) and Natural Language Generation (NLG). The RAG framework, first introduced by Lewis et al.\textcolor{blue}{\cite{lewis2020retrieval}}  combines dense retrieval methods with large scale generative models to produce responses that are both contextually relevant and factually accurate. By explicitly retrieving relevant passages from a large corpus and augmenting this information in the generation process, RAG models enhance the factual grounding of their outputs from the up-to-date knowledge.

A generic workflow of Retrieval Augmented Generation (RAG) system, showcasing how it fundamentally enhances the capabilities of Large Language Models (LLMs) by grounding their outputs in real-time, relevant information is illustrated in the Fig\textcolor{blue}{[\ref{fig:how_rag_works}]}. Unlike static models which generate responses based only on closed-world knowledge, the RAG process is structured into the following key steps:

\begin{enumerate}
    \item \textbf{Data Collection:} \\
    The workflow begins with the acquisition of relevant, domain specific textual data from various external sources, such as PDFs, structured documents, or text files. These documents represent raw data important for building a tailored knowledge base that the system will query during the retrieval process enhancing the model's ability to respond. \\

\begin{figure}[h!]
  \centering
  \includegraphics[width=1\textwidth]{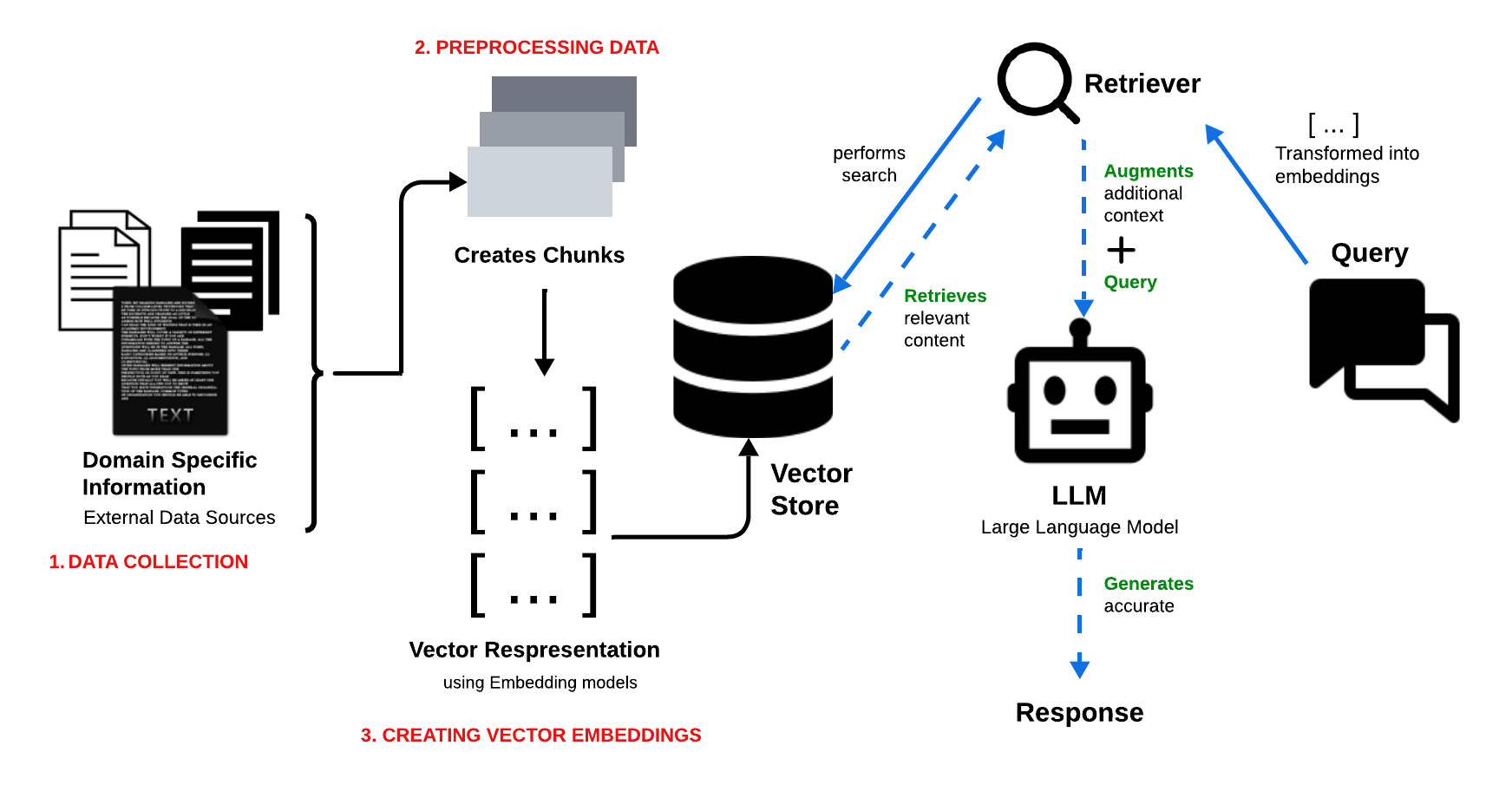}
  \caption{Architecture of Retrieval Augmented Generation(RAG) system.}
  \label{fig:how_rag_works}
\end{figure}
    
    \item \textbf{Data Preprocessing:} \\
    \label{Data Preprocessing}The collected data is then preprocessed to create manageable and meaningful chunks. Preprocessing involves cleaning the text (e.g., removing noise, formatting), normalizing it, and segmenting it into smaller units, such as \textit{tokens} (e.g., words or group of words), that can be easily indexed and retrieved later. This segmentation is necessary to ensure that the retrieval process is accurate and efficient. \\
    
    \item \textbf{Creating Vector Embeddings:} \\
    \label{Vectorization}After preprocessing, the chunks of data are transformed into \textit{vector representations} using embedding models (e.g., BERT, Sentence Transformers). These vector embeddings capture the semantic meaning of the text, allowing the system to perform similarity searches. The vector representations are stored in a \textbf{Vector Store}, an indexed database optimized for fast retrieval based on similarity measures. \\
    
    \item \textbf{Retrieval of Relevant Content:} \\
    When a \textbf{Query} is input into the system, it is first transformed into a \textit{vector embedding}, similar to the documents in the vector store. The \textbf{Retriever} component then performs a search within the vector store to identify and retrieve the most relevant chunks of information related to the query. This retrieval process ensures that the system uses the most pertinent and up-to-date information to respond to the query.\\
    
    \item \textbf{Augmentation of Context:} \\
    By merging two knowledge streams - the fixed, general knowledge embedded in the LLM and the flexible, domain-specific information augmented on demand as an additional layer of context, aligns the Large Language Model (LLM) with both established and emerging information.\\
    
    \item \textbf{Generation of Response by LLM:} \\
    The \textit{context-infused prompt}, consisting of the original user query combined with the retrieved relevant content is provided to a Large Language Model (LLM) like GPT, T5 or Llama. The LLM then processes this augmented input to generate a coherent response not only fluent but factually grounded.\\
    
    \item \textbf{Final Output:} \\
    By moving beyond the opaque outputs of traditional models, the final output of RAG systems offer several advantages: they minimize the risk of generating hallucinations or outdated information, enhance interpretability by clearly linking outputs to real-world sources, enriched with relevant and accurate responses.
    
\end{enumerate}

The RAG model framework introduces a paradigm shift in Generative AI by creating \textbf{glass-box} models. It greatly enhanced the ability of generative models to provide accurate information, especially in knowledge-intensive domains. This integration has become the backbone of many advanced NLP applications, such as chatbots, virtual assistants, and automated customer service systems.\textcolor{blue}{\cite{lewis2020retrieval}}

\subsection{When to Use RAG: Considerations for Practitioners}

Choosing between fine-tuning, using Retrieval Augmented Generation (RAG), or base models can be a challenging decision for practitioners. Each approach offers distinct advantages depending on the context and constraints of the use case. This section aims to outline the scenarios in which each method is most effective, providing a decision framework to guide practitioners in selecting the appropriate strategy. 

\subsubsection{Fine-Tuning: Domain Expertise and Customization}
Fine-tuning involves training an existing large language model (LLM) on a smaller, specialized dataset to refine its knowledge for a particular domain or task. This method excels in scenarios where accuracy, tone consistency, and deep understanding of niche contexts are essential. For instance, fine-tuning has been shown to improve a model's performance in specialized content generation, such as technical writing, customer support, and internal knowledge systems.
\\

\textbf{Advantages:} Fine-tuning embeds domain specific knowledge directly into the model, reducing the dependency on external data sources. It is particularly effective when dealing with stable data or when the model needs to adhere to a specific tone and style.
\\

\textbf{Drawbacks:} Fine-tuning is computationally expensive and often requires substantial resources for initial training. Additionally, it risks overfitting if the dataset is too narrow, making the model less generalizable.\\

\textbf{Use Case Examples:}
\begin{itemize}
    \item \textbf{Medical Diagnosis}: A fine-tuned model on medical datasets becomes highly specialized in understanding and generating medical advice based on specific terminologies and contexts.
    \item \textbf{Customer Support}: For a software company, fine-tuning on company-specific troubleshooting protocols ensures high-accuracy and consistent responses tailored to user queries.
\end{itemize}

\subsubsection{RAG: Dynamic Information and Large Knowledge Bases}
Retrieval-Augmented Generation (RAG) combines LLMs with a retrieval mechanism that allows the model to access external data sources in real-time, making it suitable for scenarios requiring up-to-date or frequently changing information. RAG systems are valuable for handling vast knowledge bases, where embedding all the information directly into the model would be impractical or impossible.
\\

\textbf{Advantages:} RAG is ideal for applications that require access to dynamic information, ensuring responses are grounded in real-time data and minimizing hallucinations. It also provides transparency, as the source of the retrieved information can be linked directly. 
\\

\textbf{Drawbacks:} RAG requires complex infrastructure, including vector databases and effective retrieval pipelines, and can be resource-intensive during inference.
\\

\textbf{Use Case Examples:}
\begin{itemize}
    \item \textbf{Financial Advisor Chatbot}: Using RAG, a chatbot can pull the latest market trends and customer-specific portfolio data to offer personalized investment advice.
    \item \textbf{Legal Document Analysis}: RAG can retrieve relevant case laws and statutes from a constantly updated database, making it suitable for legal applications where accuracy and up-to-date information are critical.
\end{itemize}

\subsubsection{When to Use Base Models}
Using base models(without fine-tuning or RAG) is appropriate when the task requires broad generalization, low-cost deployment, or rapid prototyping. Base models can handle simple use cases like generic customer support or basic question answering, where specialized or dynamic information is not required.
\\

\textbf{Advantages:} No additional training is required, making it easy to deploy and maintain. It is best for general purpose tasks or when exploring potential applications without high upfront costs. \\

\textbf{Drawbacks:} Limited performance on domain specific queries or tasks that need high levels of customization.
\begin{table}[h!]
\centering
\caption{Decision Framework for Choosing Between Fine-Tuning, RAG, and Base Models}
\begin{tabularx}{\textwidth}{|X|X|X|X|}
\hline
\textbf{Factors} & \textbf{Fine-Tuning} & \textbf{RAG} & \textbf{Base Models} \\ \hline
\textbf{Nature of the Task} & Highly specialized tasks, domain specific language & Dynamic tasks needing real time information retrieval & General tasks, prototyping, broad applicability \\ \hline
\textbf{Data Requirements} & Static or proprietary data that rarely changes & Access to up-to-date or external large knowledge bases & Does not require specialized or up-to-date information \\ \hline
\textbf{Resource Constraints} & High computational resources needed for training & Higher inference cost and infrastructure complexity & Low resource demand, quick to deploy \\ \hline
\textbf{Performance Goals} & Maximizing precision and adaptability to specific language & Providing accurate, context-aware responses from dynamic sources & Optimizing speed and cost efficiency over precision \\ \hline
\end{tabularx}
\label{tab:decision_framework}
\end{table}

To conclude, the decision framework outlined in Table[\ref{tab:decision_framework}] offers practitioners a guide to selecting the most suitable method based on their project's specific needs. Fine-Tuning is the best option for specialized, high-precision tasks with stable data; RAG should be used when access to dynamic, large-scale data is necessary; and Base Models are well-suited for general-purpose use with low resource requirements.

\subsection{Understanding the Role of PDFs in RAG}                                                                      
PDFs are paramount for RAG applications because they are widely used for distributing high-value content like research papers, legal documents, technical manuals, and financial reports, all of which contain dense, detailed information essential for training RAG models. PDFs come in various forms, allowing access to a wide range of data types—from scientific data and technical diagrams to legal terms and financial figures. This diversity makes PDFs an invaluable resource for extracting rich, contextually relevant information. Additionally, the consistent formatting of PDFs ensures accurate text extraction and context preservation, which is fundamental for generating precise responses. PDFs also include metadata (like author, keywords, and creation date) and annotations (such as highlights and comments) that provide extra context, helping RAG models prioritize sections and better understand document structure, ultimately enhancing retrieval and generation accuracy.

\subsubsection{Challenges of Working with PDFs}
\label{sec:challenges-pdf}
In RAG applications, accurate text extraction from PDFs is essential for effective retrieval and generation. However, PDFs often feature complex layouts—such as multiple columns, headers, footers, and embedded images—that complicate the extraction process. These complexities challenge RAG systems, which rely on clean, structured text for high-quality retrieval. Text extraction accuracy from PDFs decreases dramatically in documents with intricate layouts, such as multi-column formats or those with numerous figures and tables. This decline necessitates advanced extraction techniques and machine learning models tailored to diverse document structures.

Moreover, the lack of standardization in PDF creation, including different encoding methods and embedded fonts, can result in inconsistent or garbled text, further complicating extraction and degrading RAG model performance. Additionally, many PDFs are scanned documents, especially in fields like law and academia, requiring Optical Character Recognition (OCR) to convert images to text. OCR can introduce errors, particularly with low-quality scans or handwritten text, leading to inaccuracies that are problematic in RAG applications, where precise input is essential for generating relevant responses. PDFs may also contain non-textual elements like charts, tables, and images, disrupting the linear text flow required by most RAG models. Handling these elements requires specialized tools and preprocessing to ensure the extracted data is coherent and useful for RAG tasks.

\subsubsection{Key Considerations for PDF Processing in RAG Application Development}
Processing PDFs for Retrieval Augmented Generation (RAG) applications requires careful handling to ensure high-quality text extraction, effective retrieval, and accurate generation. Below are key considerations specifically tailored for PDF processing in RAG development.\\ 

\noindent
\textbf{\textit{1. Accurate Text Extraction:}} \\
Since PDFs can have complex formatting, it is essential to use reliable tools and methods to convert the PDF content into usable text for further processing.

\begin{itemize}
    \item \textbf{Appropriate Tool for Extraction:} There are tools and libraries for extracting text from PDFs for most popular programming languages (i.e: \texttt{pdfplumber} or \texttt{PyMuPDF (fitz)} for Python). These libraries handle most common PDF structures and formats, preserving the text's layout and structure as much as possible.
    \item \textbf{Verify and Clean Extracted Text:} After extracting text, always verify it for completeness and correctness. This step is essential for catching any extraction errors or artifacts from formatting.
\end{itemize}

\noindent
\textbf{\textit{2. Effective Chunking for Retrieval:}} \\
PDF documents often contain large blocks of text, which can be challenging for retrieval models to handle effectively. Chunking the text into smaller, contextually coherent pieces can improve retrieval performance.

\begin{itemize}
    \item \textbf{Semantic Chunking:} Instead of splitting text arbitrarily, use semantic chunking based on logical divisions within the text, such as paragraphs or sections. This ensures that each chunk retains its context, which is important for both retrieval accuracy and relevance.
    \item \textbf{Dynamic Chunk Sizing:} Adjust the chunk size according to the content type and the model's input limitations. For example, scientific documents might be chunked by sections, while other types of documents could use paragraphs as the primary chunking unit.
\end{itemize}

\noindent
\textbf{\textit{3. Preprocessing and Cleaning:}} \\
Preprocessing the extracted text is key for removing noise that could affect the performance of both retrieval and generative models. Proper cleaning ensures the text is consistent, relevant, and ready for further processing.

\begin{itemize}
    \item \textbf{Remove Irrelevant Content:} Use regular expressions or NLP-based rules to clean up non-relevant content like headers, footers, page numbers, and any repeating text that doesn't contribute to the document's meaning.
    \item \textbf{Normalize Text:} Standardize the text format by converting it to lowercase, removing special characters, and trimming excessive whitespace. This normalization helps create consistent input for the retrieval models.
\end{itemize}

\noindent
\textbf{\textit{4. Utilizing PDF Metadata and Annotations:}} \\
PDFs often contain metadata (such as the author, title, and creation date) and annotations that provide additional context, which can be valuable for retrieval tasks in RAG applications.

\begin{itemize}
    \item \textbf{Extract Metadata:} You can use tools specific to programming languages like \texttt{PyMuPDF} or \texttt{pdfminer.six} for Python to extract embedded metadata. This metadata can be used as features in retrieval models, adding an extra layer of context for more precise search results.
    \item \textbf{Utilize Annotations:} Extract and analyze annotations or comments within PDFs to understand important or highlighted sections. This can help prioritize content in the retrieval process.
\end{itemize}

\noindent
\textbf{\textit{5. Error Handling and Reliability:}} \\
Reliability in processing PDFs is essential for maintaining the stability and reliability of RAG applications. Implementing proper error handling and logging helps manage unexpected issues and ensures smooth operation.

\begin{itemize}
    \item \textbf{Implement Error Handling:} Use try-except blocks to manage potential errors during PDF processing. This ensures the application continues running smoothly and logs any issues for later analysis.
    \item \textbf{Use Logging for Monitoring:} Implement logging to capture detailed information about the PDF processing steps, including successes, failures, and any anomalies. This is important for debugging and optimizing the application over time.
\end{itemize}

By following these key considerations and best practices, we can effectively process PDFs for RAG applications, ensuring high-quality text extraction, retrieval, and generation. This approach ensures that your RAG models are strong, efficient, and capable of delivering meaningful insights from complex PDF documents.
%%% Local Variables:
%%% mode: latex
%%% TeX-master: "../main.tex"
%%% End:

%% file: sections/03_StudyDesign.tex
\section{Study Design}
\label{section: Study Design}

This section presents the methodology for building a Retrieval Augmented Generation (RAG) system that integrates PDF documents as a primary knowledge source. This system combines the retrieval capabilities of information retrieval (IR) techniques with the generative strengths of Large Language Models (LLMs) to produce factually accurate and contextually relevant responses, grounded in domain-specific documents.

The goal is to design and implement a RAG system that addresses the limitations of traditional LLMs, which rely solely on static, pre-trained knowledge. By incorporating real-time retrieval from domain-specific PDFs, the system aims to deliver responses that are not only contextually appropriate but also up-to-date and factually reliable.

The system begins with the collection of relevant PDFs, including research papers, legal documents, and technical manuals, forming a specialized knowledge base. Using tools and libraries, the text is extracted, cleaned, and preprocessed to remove irrelevant elements such as headers and footers. The cleaned text is then segmented into manageable chunks, ensuring efficient retrieval. These text segments are converted into vector embeddings using transformer-based models like BERT or Sentence Transformers, which capture the semantic meaning of the text. The embeddings are stored in a vector database optimized for fast similarity-based retrieval.

The RAG system architecture consists of two key components: a retriever, which converts user queries into vector embeddings to search the vector database, and a generator, which synthesizes the retrieved content into a coherent, factual response. Two types of models are considered: OpenAI’s GPT models, accessed through the Assistant API for ease of integration, and the open-source Llama model, which offers greater customization for domain-specific tasks.

In developing the system, several challenges are addressed, such as managing complex PDF layouts (e.g., multi-column formats, embedded images) and maintaining retrieval efficiency as the knowledge base grows. These challenges were highlighted during a preliminary evaluation process, where participants pointed out the difficulty of handling documents with irregular structures. Feedback from the evaluation also emphasized the need for improvements in text extraction and chunking to ensure coherent retrieval.

The design also incorporates the feedback from a diverse group of participants during a workshop session, which focused on the practical aspects of implementing RAG systems. Their input highlighted the effectiveness of the system's real-time retrieval capabilities, particularly in knowledge-intensive domains, and underscored the importance of refining the integration between retrieval and generation to enhance the transparency and reliability of the system’s outputs. This design sets the foundation for a RAG system capable of addressing the needs of domains requiring precise, up-to-date information. 

%% file: sections/04_ResultsStepbyStepGuidetoRAG.tex
\section{Results: Step-by-Step Guide to RAG}

\subsection{Setting Up the Environment}
This section walks you through the steps required to set up a development environment for Retrieval Augmented Generation (RAG) on your local machine. We will cover the installation of Python, setting up a virtual environment and configuring an IDE (VSCode).

\subsubsection{Installing Python}
If Python is not already installed on your machine, follow the steps below:

\begin{enumerate}
    \item \textbf{Download and Install Python}
    \begin{itemize}
        \item Navigate to the official Python website: \href{https://www.python.org/downloads/}{\textcolor{blue}{https://www.python.org/downloads/}}
        \item Download the latest version of Python for your operating system (Windows, macOS, or Linux).
        \item During installation, ensure that you select the option \textit{Add Python to PATH}. This is important to run Python from the terminal or command line.
        \item For Windows users, you can also: 
        \begin{itemize}
            \item Click on \textit{Customize Installation}.
            \item Select \textit{Add Python to environment variables}.
            \item Click \textit{Install Now}.\\
        \end{itemize}
    \end{itemize}
    
    \item \textbf{Verify the Installation}
    \begin{itemize}
        \item Open the terminal (Command Prompt on Windows, Terminal on macOS/Linux).
        \item Run the following command to verify that Python is installed correctly:
        \texttt{python --version}
        \item If Python is installed correctly, you should see output similar to \texttt{Python 3.x.x}.
    \end{itemize}
\end{enumerate}

\subsubsection{Setting Up an IDE}
After installing Python, the next step is to set up an Integrated Development Environment (IDE) to write and execute your Python code. We recommend Visual Studio Code (VSCode), however you are free to choose editor of your own choice. Below are the setup instructions for VSCode.

\begin{enumerate}
    \item \textbf{Download and Install VSCode}
    \begin{itemize}
        \item Visit the official VSCode website: \href{https://code.visualstudio.com/}{\textcolor{blue}{https://code.visualstudio.com/}}.
        \item Select your operating system (Windows, macOS, or Linux) and follow the instructions for installation.\\
    \end{itemize}
    
    \item \textbf{Install the Python Extension in VSCode}
    \begin{itemize}
        \item Open VSCode.
        \item Click on the \textit{Extensions} tab on the left-hand side (it looks like a square with four pieces).
        \item In the Extensions Marketplace, search for \textit{Python}.
        \item Install the Python extension by Microsoft. This will allow VSCode to support Python code.
    \end{itemize}
\end{enumerate}

\subsubsection{Setting Up a Virtual Environment}
\label{sec:4.3.2 Setting up virtual environment}
A virtual environment allows you to install libraries and dependencies specific to your project without affecting other projects on your machine.

\begin{enumerate}
    \item \textbf{Open the Terminal in VSCode}
    \begin{itemize}
        \item Press \texttt{Ctrl + `} (or \texttt{Cmd + `} on Mac) to open the terminal in VSCode.
        \item Alternatively, navigate to \textit{View - Terminal} in the menu.
    \end{itemize}
    \begin{itemize}  
        \item In the terminal, use the \texttt{mkdir} command to create a new folder for your project. For example, to create a folder named \texttt{my-new-project}, type: \begin{verbatim} mkdir my-new-project \end{verbatim} 
    \end{itemize}
    \begin{itemize}
        \item Use the \texttt{cd} command to change directories and navigate to the folder where your project is located. For example: \begin{verbatim} cd path/to/your/project/folder/my-new-project \end{verbatim}
    \end{itemize}

    \item \textbf{Create a Virtual Environment}    
    \begin{itemize}
        \item For Windows, run the following commands:
        \begin{verbatim}
        python -m venv my_rag_env
        my_rag_env\Scripts\activate
        \end{verbatim}

        \item For Mac/Linux, run the following commands:
        \begin{verbatim}
        python3 -m venv my_rag_env
        source my_rag_env/bin/activate
        \end{verbatim}
    \end{itemize}
    
    \item \textbf{Configure VSCode to Use the Virtual Environment}
    \begin{itemize}
        \item Open the Command Palette by pressing \texttt{Ctrl + Shift + P} (or \texttt{Cmd + Shift + P} on Mac).
        \item Type \textit{Python: Select Interpreter} in the Command Palette.
        \item Select your virtual environment, \texttt{my\_rag\_env}, from the list.
    \end{itemize}

\end{enumerate}

With your virtual environment now configured, you are ready to install project specific dependencies and manage Python packages independently for each approach. This setup allows you to create separate virtual environments for the two approaches outlined in Sections{\textcolor{blue}{[\ref{OpenAI's Assistant API}][\ref{Open-Source Llama}]}. By isolating your dependencies, you can ensure that the OpenAI Assistant API-based{\textcolor{blue}{[\ref{OpenAI's Assistant API}]} and Llama-based {\textcolor{blue}{[\ref{Open-Source Llama}]} Retrieval Augmented Generation (RAG) systems are developed and managed in their respective environments without conflicts or dependency issues. This practice also helps maintain cleaner, more manageable development workflows for both models, ensuring that each approach functions optimally with its specific requirements.

\subsection{Two Approaches to RAG: Proprietary and Open source}

This section introduces a structured guide for developing Retrieval Augmented Generation (RAG) systems, focusing on two distinct approaches: \textit{using OpenAI's Assistant API (GPT Series)} and an \textit{open-source Large Language Model (LLM) Llama} and thus divided into two subsections{\textcolor{blue}{[\ref{OpenAI's Assistant API}][\ref{Open-Source Llama}]}. The objective is to equip developers with the knowledge and practical steps necessary to implement RAG systems effectively, while highlighting common mistakes and best practices at each stage of the process. Each subsection is designed to provide practical insights into setup, development, integration, customization and optimization to generate well-grounded and aligned outputs.

\begin{table}[ht]
\centering
\renewcommand{\arraystretch}{1.5} % Increase row height
\setlength{\tabcolsep}{5pt} % Adjust column separation
\caption{Comparison of RAG Approaches: OpenAI vs. Llama}
\label{tab:comparison_rag_approaches}
\begin{tabularx}{\textwidth}{|p{3cm}|X|X|}
\hline
\textbf{Feature} & \textbf{OpenAI's Assistant API (GPT Series)} & \textbf{Llama (Open-Source LLM Model)} \\
\hline
\textbf{Ease of Use} & High. Simple API calls with no model management & Moderate. Requires setup and model management \\
\hline
\textbf{Customization} & Limited to prompt engineering and few-shot learning & High. Full access to model fine-tuning and adaptation \\
\hline
\textbf{Cost} & Pay-per-use pricing model & Upfront infrastructure costs; no API fees \\
\hline
\textbf{Deployment Flexibility} & Cloud-based; depends on OpenAI’s infrastructure & Highly flexible; can be deployed locally or in any cloud environment \\
\hline
\textbf{Performance} & Excellent for a wide range of general NLP tasks & Excellent, particularly when fine-tuned for specific domains \\
\hline
\textbf{Security and Data Privacy} & Data is processed on OpenAI servers; privacy concerns may arise & Full control over data and model; suitable for sensitive applications \\
\hline
\textbf{Support and Maintenance} & Strong support, documentation, and updates from OpenAI & Community-driven; updates and support depend on community efforts \\
\hline
\textbf{Scalability} & Scalable through OpenAI's cloud infrastructure & Scalable depending on infrastructure setup \\
\hline
\textbf{Control Over Updates} & Limited; depends on OpenAI’s release cycle & Full control; users can decide when and how to update or modify the model \\
\hline
\end{tabularx}
\end{table}

In addition to the two primary approaches discussed in this guide there are several alternative frameworks and methodologies for developing Retrieval Augmented Generation (RAG) systems. Each of these options such as Cohere, AI21’s Jurassic-2, Google’s PaLM, and Meta’s OPT have their merits and trade-offs in terms of deployment flexibility, cost, ease of use, and performance.

We have selected \textbf{OpenAI's Assistant API (GPT Series)} and \textbf{Llama} for this guide based on their wide adoption, proven capabilities, and distinct strengths in developing RAG systems. As highlighted in comparison Table{\textcolor{blue}{[\ref{tab:comparison_rag_approaches}]} \textit{OpenAI’s Assistant API} provides a simple and developer-friendly black-box, allowing quick integration and deployment without the need for extensive model management or infrastructure setup with high quality outputs. In contrast, as an open-source model, Llama allows developers to have full control over the model’s architecture, training data, and fine-tuning process, allowing for precise customization to suit specific requirements such as demand control, flexibility, and cost-efficiency. This combination makes these two options highly valuable for diverse RAG system development needs.
\newpage
\subsubsection{Using OpenAI's Assistant API : GPT Series}
\label{OpenAI's Assistant API}

While the OpenAI Completion API is effective for simple text generation tasks, the Assistant API is a superior choice for developing RAG systems. The Assistant API supports \textbf{multi-modal operations} (such as text, images, audio, and video inputs) by combining text generation with file searches, code execution, and API calls. For a RAG system, this means an assistant can retrieve documents, generate vector embeddings, search for relevant content, augment user queries with additional context, and generate responses—all in a seamless, integrated workflow. It includes memory management across sessions, so the assistant \textit{remembers} past queries, retrieved documents, or instructions. Assistants can be configured with specialized instructions, behaviors, parameters other than custom tools that makes this API far more powerful for developing RAG systems.

\begin{figure}[h!]
  \centering
  \includegraphics[width=1\textwidth]{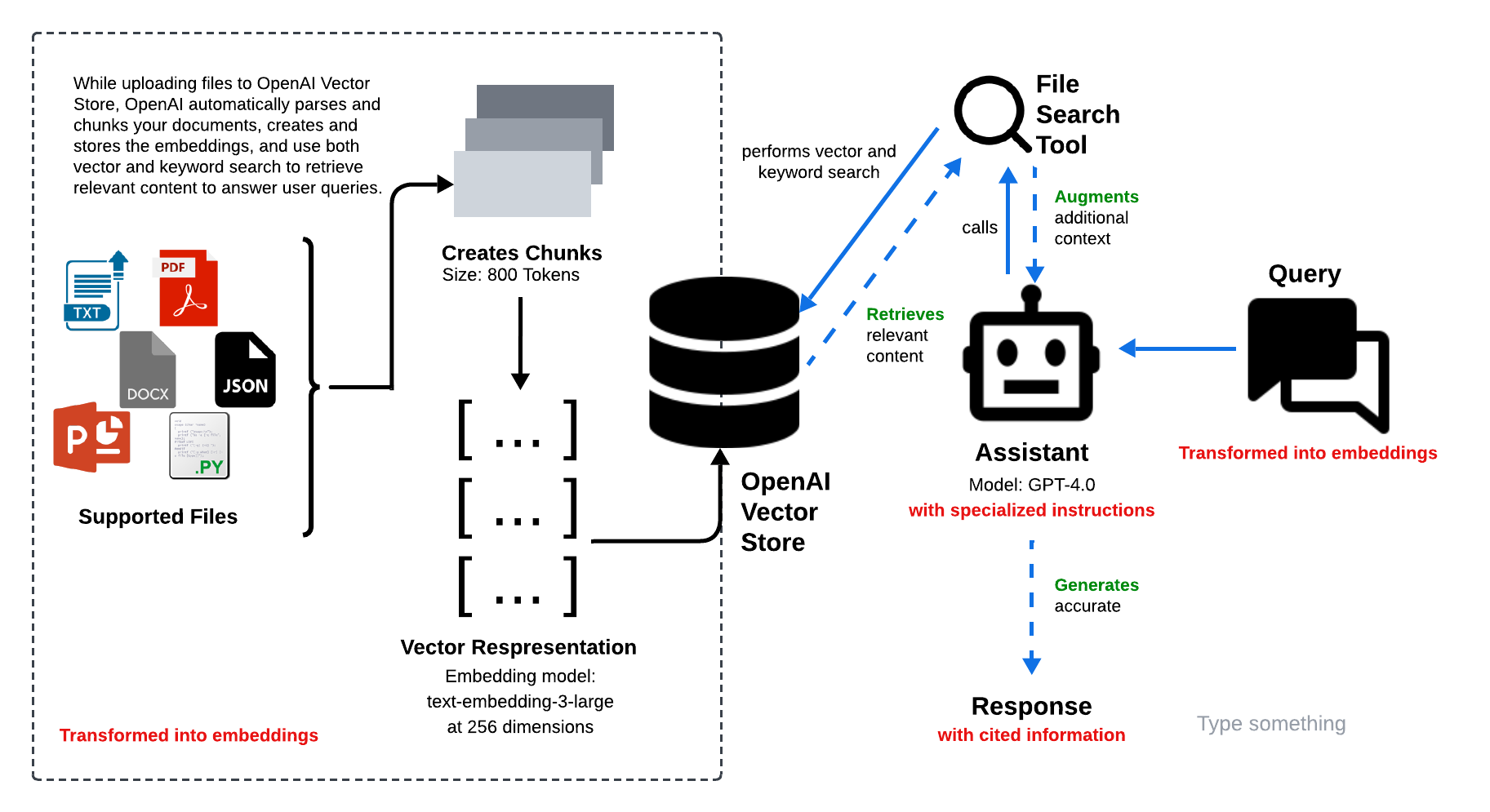}
  \caption{Open AI's Assistant API Workflow}
  \label{fig:rag_with_AssistantAPI}
\end{figure}

This subsection provides a step-by-step guide and code snippets to utilize the OpenAI's \textbf{File Search tool} within the Assistant API, as illustrated in Fig{\textcolor{blue}{[\ref{fig:rag_with_AssistantAPI}]} to implement RAG. The diagram shows how after the domain specific data ingestion of supported files (such as PDFs, DOCX, JSON, etc.), the data prepocessing{\textcolor{blue}{[\ref{Data Preprocessing}]} and vectorization{\textcolor{blue}{[\ref{Vectorization}]} is handled by Assistant API. These vectors are stored in OpenAI \textbf{Vector Store}, which the File Search tool can query to retrieve relevant content. The assistant then augments the context and generates accurate responses based on specialized instructions and the retrieved information. This integrated process is covered in detailed steps below:\\

\begin{enumerate}
\item \textbf{Environment Setup and API Configuration} \\

Setting up your environment and configuring access to OpenAI's API is the foundational step. \\

\begin{enumerate}
\item
    \href{https://www.google.com/url?sa=t&rct=j&q=&esrc=s&source=web&cd=&ved=2ahUKEwicmZjjyLaIAxVASFUIHdcmJVQQFnoECBkQAQ&url=https%3A%2F%2Fplatform.openai.com%2Fsignup%2F&usg=AOvVaw1TS2xAOb51_MsLhOBmbqTM&opi=89978449}{Create an OpenAI Account}.
\item Once logged in, navigate to the \href{https://platform.openai.com/account/api-keys}{OpenAI API dashboard}. Generate a New Project API Key. 
\item Depending on your usage and plan, OpenAI may require you to set up billing information. Navigate to the \href{https://platform.openai.com/settings/organization/billing/overview}{Billing} section in the dashboard to add your payment details. Refer to Appendix{\textcolor{blue}{[\ref{Appendix}]}} for cost estimations.\\
\end{enumerate}

Store your API key securely. A \textbf{.env} file is used to securely store environment variables, such as your OpenAI API key. \\

\begin{enumerate}
\item \textbf{Set Up a New Project Folder and Virtual Environment:} First, create a new folder for your project. Ensure that a virtual environment is already set up in this folder, as described in section{\textcolor{blue}{[\ref{sec:4.3.2 Setting up virtual environment}]}}. 
\item
\textbf{Create a .env File:} Inside your new folder, make a file called \textit{.env}. This file will store your OpenAI API key.
\item\textbf{Add Your API Key:} Open the \textit{.env} file and paste your OpenAI API key in this format:
\begin{lstlisting}
OPENAI_API_KEY=your_openai_api_key_here
\end{lstlisting}
\textit{Be sure to replace with your actual API key.}
\item\textbf{Save the .env File:} After adding your key, save the .env file in the same folder where you'll keep your Python files.
\item\textbf{Install Necessary Python Packages:} To make everything work, you need two tools: openai and python-dotenv. Open a terminal (or Command Prompt) and run this command to install them:
\begin{lstlisting}
pip install python-dotenv openai
\end{lstlisting} 
\textit{If you need specific version of these tools used for the code in GitHub repository, you can install them like this:} 
\begin{lstlisting}
pip install python-dotenv==1.0.1 openai==1.37.2
\end{lstlisting} 
\item\textbf{Create the Main Python File:} In the same folder, create a new file called \textbf{main.py}. All the code snippets attached in this entire section[\ref{OpenAI's Assistant API}] should be implemented within this file.
\end{enumerate}

To interact with the OpenAI API and load environment variables, you need to import the necessary libraries. The \textit{dotenv} library will be used to load environment variables from the \textit{.env} file.\\

\textbf{Code Example: Import Dependencies }
\begin{lstlisting}
import os
import openai
import time
from dotenv import load_dotenv
\end{lstlisting}

Next, you need to load the environment variables from the .env file and set up the OpenAI API \textbf{client}. This is important for authenticating your requests to the OpenAI service and setting up the connection to interact with OpenAI's Assistant API.\\

\noindent
\textbf{Code Example: Set OpenAI API Key and LLM }
\begin{lstlisting}
# Load environment variables from .env file
load_dotenv()

# Check if OPENAI_API_KEY is set
openai_api_key = os.getenv("OPENAI_API_KEY")
if not openai_api_key:
    raise EnvironmentError("Error: OPENAI_API_KEY is not set in the environment. Please set it in the .env file.")

# Set OpenAI key and model
openai.api_key = openai_api_key
client = openai.OpenAI(api_key=openai.api_key)
model_name = "gpt-4o"  # Any model from GPT series
\end{lstlisting}
\vspace{1em}

\item \textbf{Understanding the Problem Domain and Data Requirements} \\

To develop an effective solution for managing and retrieving information, it's important to understand the problem domain and identify the specific data requirements and \textit{not just provide any data}. For a deeper insight into the challenges of handling PDFs, refer to Section{\textcolor{blue}{[\ref{sec:challenges-pdf}]}. Given that this paper focuses on working with PDFs, it is important to emphasize the significance of having relevant and clean data within these documents.\\  

\textbf{Organize the Knowledge Base Files:} After selecting the PDF(s) for your external knowledge base, create a folder named \textbf{Upload} in the project directory, and place all the selected PDFs inside this folder.

\begin{tcolorbox}[title=Common Mistakes and Best Practices, colframe=gray, colback=gray!10, fonttitle=\bfseries]
\begin{tabularx}{\textwidth}{X} % 'X' column type for automatic line breaking
    \textbf{Mistake: Irrelevant or inconsistent data} Poor-structured data in PDFs downgrades the quality of embeddings generated for Large Language Models (LLMs), hindering them to understand and process the content more accurately.\\    
    \textbf{Best Practice: Ensure data consistency and relevance} \ All PDFs uploaded to the vector store are consistent in format and highly relevant to the problem domain. \\
    \textbf{Best Practice: Use descriptive file names and metadata} \ Using descriptive file names and adding relevant metadata can help with debugging, maintenance, and retrieval tasks. Files should be named in a way that reflects their content or relevance to the RAG system. \\
\end{tabularx}
\end{tcolorbox}
\vspace{1em}
The following Python code defines a function to upload multiple PDF files from a specified directory to OpenAI vector store, which is a common data structure used for storing and querying high-dimensional vectors, often for machine learning and AI applications. It ensures the directory and files are valid before proceeding with the upload and collects and returns the uploaded files' IDs.\\

\textcolor{red}{\textit{NOTE: The function is called to run only when a new vector store is created, meaning it won't upload additional files to an existing vector store. You can modify the logic as needed to suit your requirements. \\ Also, please be aware that the files are stored on external servers, such as OpenAI’s infrastructure. OpenAI has specific policies regarding data access and usage to protect user privacy and data security. They state that they do not use customer data to train their models unless explicitly permitted by the user. For more details refer:\textcolor{blue}{\href{https://openai.com/policies/privacy-policy/}{https://openai.com/policies/privacy-policy/}}. Additionally, the data stored can be deleted easily when necessary either via code:\textcolor{blue}{\href{https://platform.openai.com/docs/api-reference/files/delete}{https://platform.openai.com/docs/api-reference/files/delete}} or from the user interface by clicking the delete button here:\\
\textcolor{blue}{\href{https://platform.openai.com/storage/files/}{https://platform.openai.com/storage/files/}}. }}\\

\textbf{Code Example: Upload PDF(s) to the OpenAI Vector Store}
\begin{lstlisting}
def upload_pdfs_to_vector_store(client, vector_store_id, directory_path):
    try:
        if not os.path.exists(directory_path):
            raise FileNotFoundError(f"Error: Directory '{directory_path}' does not exist.")
        if not os.listdir(directory_path):
            raise ValueError(f"Error: Directory '{directory_path}' is empty. No files to upload.")
        
        file_ids = {}
        # Get all PDF file paths from the directory
        file_paths = [os.path.join(directory_path, file) for file in os.listdir(directory_path) if file.endswith(".pdf")]

        # Check if there are any PDFs to upload
        if not file_paths:
            raise ValueError(f"Error: No PDF files found in directory '{directory_path}'.")

        # Iterate through each file and upload to vector store
        for file_path in file_paths:
            file_name = os.path.basename(file_path)
            
            # Upload the new file
            with open(file_path, "rb") as file:
                uploaded_file = client.beta.vector_stores.files.upload(vector_store_id=vector_store_id, file=file)
                print(f"Uploaded file: {file_name} with ID: {uploaded_file.id}")
                file_ids[file_name] = uploaded_file.id

        print(f"All files have been successfully uploaded to vector store with ID: {vector_store_id}")
        return file_ids

    except Exception as e:
        print(f"Error uploading files to vector store: {e}")
        return None
\end{lstlisting}
\vspace{1em}
\item \textbf{Creating and Managing Vector Stores in OpenAI} \\

OpenAI Vector stores are used to store files for use by the \textit{file search} tool in Assistant API. This step involves initializing a vector store for storing vector embeddings of documents and retrieving them when needed.\\

\textbf{Code Example: Initialize a Vector Store for RAG}
\begin{lstlisting}
# Get/Create Vector Store
def get_or_create_vector_store(client, vector_store_name):
    if not vector_store_name:
        raise ValueError("Error: 'vector_store_name' is not set. Please provide a valid vector store name.")
    
    try:
        # List all existing vector stores
        vector_stores = client.beta.vector_stores.list()
        
        # Check if the vector store with the given name already exists
        for vector_store in vector_stores.data:
            if vector_store.name == vector_store_name:
                print(f"Vector Store '{vector_store_name}' already exists with ID: {vector_store.id}")
                return vector_store
        
        # Create a new vector store if it doesn't exist
        vector_store = client.beta.vector_stores.create(name=vector_store_name)
        print(f"New vector store '{vector_store_name}' created with ID: {vector_store.id}")
        
        # Upload PDFs to the newly created vector store (assuming 'Upload' is the directory containing PDFs)
        upload_pdfs_to_vector_store(client, vector_store.id, 'Upload')
        return vector_store

    except Exception as e:
        print(f"Error creating or retrieving vector store: {e}")
        return None
\end{lstlisting}

\begin{tcolorbox}[title=Common Mistakes and Best Practices, colframe=gray, colback=gray!10, fonttitle=\bfseries]
\begin{tabularx}{\textwidth}{X} % 'X' column type for automatic line breaking
    \textbf{Mistake: Ignoring context and query augmentation strategies} \ Relying solely on the vector embeddings without considering query-specific context or augmentation can lead to suboptimal responses. \\
    \textbf{Best Practice: Augment Queries with Contextual Information} \ Incorporate additional contextual information when forming queries to improve retrieval quality. Using techniques like relevance feedback or pseudo-relevance feedback can help refine search results.[\cite{Li_2023}][\cite{Arampatzis_2021}] \\
    \textbf{Best Practice: Handle Naming Conflicts Gracefully} \ When creating vector stores, consider adding a timestamp or unique identifier to the vector store name to avoid naming conflicts and make it easier to manage multiple vector stores.\\    
    \textbf{Best Practice: Chunking strategy} \ By default OpenAI uses a \textit{max chunk size tokens of 800} and \textit{chunk overlap tokens of 400} to chunk the file(s) for Vector Stores. Properly sized chunks ensure that each chunk contains a coherent and contextually meaningful piece of information. If chunks are too large, they may contain unrelated content, conversely, if chunks are too small, they may lack sufficient context to be useful. Configure the variables accordingly the PDF(s).
\end{tabularx}
\end{tcolorbox}
\vspace{1em}
Once the functions to upload PDF file(s) and creating a vector store are defined you can call it to create \textbf{Knowledge Base} for your project by providing \textbf{vector store name} and store in a \textit{vector store} object as shown below: \\

\textbf{Code Example: Creating Vector Store Object}
\begin{lstlisting}
vector_store_name = ""  # Ensure this is set to a valid name
vector_store = get_or_create_vector_store(client, vector_store_name)
\end{lstlisting}

\vspace{1em}
\item \textbf{Creating Assistant with Specialized Instructions} \\

After setting up the vector store, the next step is to create an AI assistant using the OpenAI API. This assistant will be configured with specialized instructions and tools to perform RAG tasks effectively. Set the \textbf{assistant name}, \textbf{description} and \textbf{instructions} properties accordingly. Refer to the best practices, if needed you can also play with the \textbf{temperature} and \textbf{top p} values as per the project needs for random or deterministic responses. 

\textbf{Code Example: Create and Configure Assistant}
\begin{lstlisting}
# Get/Create Assistant
def get_or_create_assistant(client, model_name, vector_store_id):

    assistant_name = ""  # Ensure this is set to a valid name
    description = ""  # Ensure Purpose of Assistant is set here
    instructions = ""  # Ensure Specialized Instructions for Assistant and Conversation Structure is set here)
    
    try:
        assistants = client.beta.assistants.list()
        for assistant in assistants.data:
            if assistant.name == assistant_name:
                print("AI Assistant already exists with ID:" + assistant.id)
                return assistant

        assistant = client.beta.assistants.create(
            model=model_name,
            name=assistant_name,
            description=description,  
            instructions=instructions,  
            tools=[{"type": "file_search"}],
            tool_resources={"file_search": {"vector_store_ids": [vector_store_id]}},
            temperature=0.7,  # Temperature for sampling
            top_p=0.9  # Nucleus sampling parameter
        )
        print("New AI Assistant created with ID:" + assistant.id)
        return assistant

    except Exception as e:
        print(f"Error creating or retrieving assistant: {e}")
        return None

assistant = get_or_create_assistant(client, model_name, vector_store.id)
\end{lstlisting}
\begin{tcolorbox}[title=Common Mistakes and Best Practices, colframe=gray, colback=gray!10, fonttitle=\bfseries]
\begin{tabularx}{\textwidth}{X} % 'X' column type for automatic line breaking   
    \textbf{Best Practice: Ask the Model to adopt a Persona} \ Provide specialized context-Rich \textbf{\textit{instructions}} that guide the assistant on how to handle queries, what tone to use (e.g., formal, friendly), and which domains to prioritize. This ensures the assistant generates more accurate and contextually appropriate responses. \\
    \textbf{Best Practice: Use inner monologue or Conversation Structure} \ The idea of inner monologue as a part of \textbf{\textit{instructions}} is to instruct the model to put parts of the output into a structured format. This help understand the reasoning process that a model uses to arrive at a final answer. \\   
    \textbf{Best Practice: Fine-Tune Model Parameters Based on Use Case} \ Adjust parameters such as \textbf{\textit{temperature }}(controls randomness) and \textbf{\textit{top p}} (controls diversity) based on the application needs. This can impact the coherence and creativity of the assistant's outputs. For a customer support assistant, a lower temperature may be preferable for consistent responses, while a more creative application might benefit from a higher temperature. \\
    \textbf{Best Practice: Classifying Queries into Categories} \ For tasks in which lots of independent sets of instructions are needed to handle different cases, it can be beneficial to first classify the type of query and to use that classification to determine which instructions are needed. \\
\end{tabularx}
\end{tcolorbox}
\vspace{1em}
\item \textbf{Creating Conversation Thread} \\

Creating a thread, initializes a context-aware conversation session where the AI assistant can interact with the user, retrieve relevant information from the vector store, and generate responses based on that context. Additionally, \textbf{\textit{tool resources}} can be attached to the Assistant API threads that are made available to the assistant's tools in this thread. \\

\textit{This capability is essentially important when there is a need to use the same AI assistant with different tools for different threads. They can be dynamically managed to suit the requirements for topic-specific threads, reusing the same Assistant across different contexts or overwriting assistant tools for a specific thread. } \\

\textbf{Code Example: Initialize a thread for conversation}
\begin{lstlisting}
# Create thread
thread_conversation = {
    "tool_resources": {
        "file_search": {
            "vector_store_ids": [vector_store.id]
        }
    }
}

message_thread = client.beta.threads.create(**thread_conversation)
\end{lstlisting}

\vspace{1em}
\item \textbf{Initiating a Run} \\

A \textbf{\textit{Run}} represents an execution on a thread.This step involves sending user input to the assistant, which then processes it using the associated resources, retrieves information as needed, and returns a response that could include dynamically fetched citations or data from relevant documents.\\

This following code allows a user to ask questions to an assistant in a loop. It sends the user's question, waits for the assistant to think and respond, and then displays the response word by word. The process repeats until the user types "exit" to quit.\\

\textbf{Code Example: Interact with the LLM}
\begin{lstlisting}
# Interact with assistant
while True:
    user_input = input("Enter your question (or type 'exit' to quit): ")
    
    if user_input.lower() == 'exit':
        print("Exiting the conversation. Goodbye!")
        break

    # Add a message to the thread with the new user input
    message_conversation = {
        "role": "user",
        "content": [
            {
                "type": "text",
                "text": user_input
            }
        ]
    }

    message_response = client.beta.threads.messages.create(thread_id=message_thread.id, **message_conversation)

    run = client.beta.threads.runs.create(
        thread_id=message_thread.id,
        assistant_id=assistant.id
    )

    response_text = ""
    citations = []
    processed_message_ids = set()  

    while True:
        run_status = client.beta.threads.runs.retrieve(run.id, thread_id=message_thread.id)
        if run_status.status == 'completed':
            break
        elif run_status.status == 'failed':
            raise Exception(f"Run failed: {run_status.error}")
        time.sleep(1)  

    while True:
        response_messages = client.beta.threads.messages.list(thread_id=message_thread.id)

        new_messages = [msg for msg in response_messages.data if msg.id not in processed_message_ids]
        
        for message in new_messages:
            if message.role == "assistant" and message.content:
                message_content = message.content[0].text
                annotations = message_content.annotations

                for index, annotation in enumerate(annotations):
                    message_content.value = message_content.value.replace(annotation.text, f"[{index}]")
                    if file_citation := getattr(annotation, "file_citation", None):
                        cited_file = client.files.retrieve(file_citation.file_id)
                        citations.append(f"[{index}] {cited_file.filename}")
                
                words = message_content.value.split()
                for word in words:
                    print(word, end=' ', flush=True)
                    time.sleep(0.05)
                
                processed_message_ids.add(message.id)

        if any(msg.role == "assistant" and msg.content for msg in new_messages):
            break

        time.sleep(1)

    if citations:
        print("\nSources:", ", ".join(citations))
    print("\n") 
\end{lstlisting}}

\end{enumerate}

\noindent
The flexibility of configuring the assistant and thread or assistant-level tools OpenAI's API makes this approach highly versatile for various use cases. By following the steps outlined in this section and adhering to the provided best practices, developers can effectively build a powerful RAG system.

\subsubsection{Using Open-Source LLM Model: Llama}
\label{Open-Source Llama}
We will utilize Ollama, an open-source framework that implements the Llama model, to incorporate Llama-based question generation capabilities within our application. By employing Ollama, we can process user input and generate contextually relevant questions directly through the terminal, offering an efficient and scalable solution for natural language processing tasks in a local environment. This integration will enable seamless question generation without relying on external API services, ensuring both privacy and computational efficiency.

\begin{enumerate}
\item \textbf{Install the following Python libraries}
\begin{lstlisting}[language=bash]
pip install pymupdf langchain-huggingface faiss-cpu
pip install oLlama sentence-transformers sentencepiece 
pip install langchain-community
\end{lstlisting}

\vspace{10pt} % Adds 10pt of space between the items

\item \textbf{Converting PDFs to Text Files}

Save this script as \texttt{pdf\_to\_text.py}. This script converts PDF files in a given folder into text files. You have to create one folder at the same directory. The folder name should be \textit{Data}. You have to keep your PDF files in this \textit{Data} folder for the further process. \href{https://github.com/GPT-Laboratory/RAG-LLM-Development-Guidebook-from-PDFs/blob/main/RAGUsingLlama3.1/pdf_to_text.py}{Check the GitHub Link of the Code}.\\

\textbf{Code Example:}
\begin{lstlisting}[language=Python]
import os
import fitz  # PyMuPDF for reading PDFs

def convert_pdfs_to_text(pdf_folder, text_folder):
    if not os.path.exists(text_folder):
        os.makedirs(text_folder)

    for file_name in os.listdir(pdf_folder):
        if file_name.endswith(".pdf"):
            file_path = os.path.join(pdf_folder, file_name)
            text_file_name = os.path.splitext(file_name)[0] + ".txt"
            text_file_path = os.path.join(text_folder, text_file_name)

            with fitz.open(file_path) as doc:
                text = ""
                for page in doc:
                    text += page.get_text()

            with open(text_file_path, "w", encoding="utf-8") as text_file:
                text_file.write(text)
            print(f"Converted {file_name} to {text_file_name}")

if __name__ == "__main__":
    pdf_folder = "Data"
    text_folder = "DataTxt"
    convert_pdfs_to_text(pdf_folder, text_folder)
\end{lstlisting}
\vspace{1em}
\textbf{Functionality}: Converts all PDFs in the \texttt{Data} folder to text files and saves them in the \texttt{DataTxt} folder.\\

\item \textbf{Creating the FAISS Index}

Save this script as \texttt{txt\_to\_index.py}. This script generates a FAISS index from the text files in the \texttt{DataTxt} folder. \href{https://github.com/GPT-Laboratory/RAG-LLM-Development-Guidebook-from-PDFs/blob/main/RAGUsingLlama3.1/txt_to_index.py}{Check the GitHub link for the Code.}\\

\textbf{Code Example:}
\begin{lstlisting}[language=Python]
import os
from langchain_huggingface import HuggingFaceEmbeddings
from langchain_community.vectorstores import FAISS

def load_text_files(text_folder):
    texts = []
    for file_name in os.listdir(text_folder):
        if file_name.endswith(".txt"):
            file_path = os.path.join(text_folder, file_name)
            with open(file_path, "r", encoding="utf-8") as file:
                texts.append(file.read())
    return texts

def create_faiss_index(text_folder, index_path, embedding_model='sentence-transformers/all-MiniLM-L6-v2'):
    texts = load_text_files(text_folder)
    embeddings = HuggingFaceEmbeddings(model_name=embedding_model)
    vector_store = FAISS.from_texts(texts, embeddings)

    vector_store.save_local(index_path)
    print(f"FAISS index saved to {index_path}")

if __name__ == "__main__":
    text_folder = "DataTxt"
    index_path = "DataIndex"
    create_faiss_index(text_folder, index_path)
\end{lstlisting}
\vspace{1em}
\textbf{Functionality}: Creates a FAISS index and saves it in the \texttt{DataIndex/} folder.\\
Here, \texttt{sentence-transformers/all-MiniLM-L6-v2} is a compact, fast transformer model that generates sentence embeddings for tasks like semantic search and text similarity. It's efficient and ideal for quick, accurate text retrieval in applications like FAISS indexing.\\

\item \textbf{Setting Up OLlama and Llama 3.1}

Before implementing the last script that handles user queries and generates responses using a Large Language Model (LLM), we need to select an open-source LLM. By using OLlama as the model runner, we can easily integrate Llama 3.1 into our system.

\vspace{10pt}

\begin{enumerate}
    \item \textbf{Step 1: Download OLlama}
    
    To get started with OLlama, follow these steps:
    \begin{enumerate}
        \item Visit the official \href{https://oLlama.com/download}{OLlama website}.
        \item Choose the version suitable for your operating system (Windows, macOS, or Linux).
        \item Download and install the appropriate installer from the website.
    \end{enumerate}

    \vspace{10pt}

    \item \textbf{Step 2: Install Llama 3.1 Using OLlama}
    
    After installing OLlama, you can download and install Llama 3.1 by running the following command in your PowerShell or CMD terminal:
    \begin{lstlisting}[language=bash]
oLlama pull Llama3.1
    \end{lstlisting}
    
    \vspace{10pt}

    \item \textbf{Step 3: Running Llama 3.1}

    Once the Llama 3.1 model is installed, you can start using it by running a command similar to this:
    \begin{lstlisting}[language=bash]
oLlama run Llama3.1
    \end{lstlisting}
    This command runs the Llama 3.1 model, allowing you to ask questions directly in the terminal and interact with the model in real time.

    \vspace{10pt}

    \item \textbf{Step 4: Test OLlama in VS Code}

    create a \texttt{.bat} file to avoid typing the full path each time. Open Notepad, add this:
    \begin{lstlisting}[language=bash]
@echo off
"C:\path\to\OLlama\oLlama.exe" %*
    \end{lstlisting}
    
    Replace the path with your own. Save the file as \texttt{oLlama.bat} directly in your project folder.\\
    In the VS Code terminal, run the \texttt{.bat} file with the command:
    
    \begin{lstlisting}[language=bash]
.\oLlama.bat run Llama3.1
\end{lstlisting}
\end{enumerate}

\vspace{10pt}

\item \textbf{Implementing RAG-Based Question Generation}

Save this script as \texttt{main.py}. This script retrieves relevant documents using FAISS and generates questions based on the retrieved context using OLlama and Llama 3.1. \href{https://github.com/GPT-Laboratory/RAG-LLM-Development-Guidebook-from-PDFs/blob/main/RAGUsingLlama3.1/main.py}{Check the GitHub Code}.\\

\textbf{Code Example:}
\begin{lstlisting}[language=Python]
import os
from langchain_community.vectorstores import FAISS
from langchain_huggingface import HuggingFaceEmbeddings
from langchain.prompts import PromptTemplate
from langchain.chains import RetrievalQA
from langchain_community.llms import OLlama

# Load FAISS index
def load_faiss_index(index_path, embedding_model):
    # Load the FAISS index using the same embedding model
    embeddings = HuggingFaceEmbeddings(model_name=embedding_model)
    vector_store = FAISS.load_local(index_path, embeddings, allow_dangerous_deserialization=True)
    return vector_store

# Create the RAG system using FAISS and OLlama (Llama 3.1)
def create_rag_system(index_path, embedding_model='sentence-transformers/all-MiniLM-L6-v2', model_name="Llama3.1"):
    # Load the FAISS index
    vector_store = load_faiss_index(index_path, embedding_model)

    # Initialize the OLlama model (Llama3.1)
    llm = OLlama(model=model_name)

    # Create a more detailed prompt template
    prompt_template = """
    You are an expert assistant with access to the following context extracted from documents. Your job is to answer the user's question as accurately as possible, using the context below.

    Context:
    {context}

    Given this information, please provide a comprehensive and relevant answer to the following question:
    Question: {question}

    If the context does not contain enough information, clearly state that the information is not available in the context provided.
    If possible, provide a step-by-step explanation and highlight key details.
    """

    # Create a template for formatting the input for the model
    prompt = PromptTemplate(
        input_variables=["context", "question"],
        template=prompt_template
    )

    # Create a RetrievalQA chain that combines the vector store with the model
    qa_chain = RetrievalQA.from_chain_type(
        llm=llm,
        chain_type="stuff",
        retriever=vector_store.as_retriever(),
        chain_type_kwargs={"prompt": prompt}
    )

    return qa_chain

# Function to run the RAG system with a user question
def get_answer(question, qa_chain):
    answer = qa_chain.run(question)
    return answer

if __name__ == "__main__":
    # Path to the FAISS index directory
    index_path = "DataIndex"

    # Initialize the RAG system
    rag_system = create_rag_system(index_path)

    # Get user input and generate the answer
    while True:
        user_question = input("Ask your question (or type 'exit' to quit): ")
        if user_question.lower() == "exit":
            print("Exiting the RAG system.")
            break
        answer = get_answer(user_question, rag_system)
        print(f"Answer: {answer}")
\end{lstlisting}
\vspace{1em}
\textbf{Functionality}: The script takes a user query from the terminal. It retrieves relevant documents using FAISS. Then it generates a answer using the retrieved context with OLlama and Llama 3.1.\\

\begin{tcolorbox}[title=Common Mistakes and Best Practices, colframe=gray, colback=gray!10, fonttitle=\bfseries]
\begin{tabularx}{\textwidth}{X} % 'X' column type for automatic line breaking   
    \textbf{Incompatible embeddings} \ The FAISS index is typically created using a specific embeddings model. If a different embeddings model is used during querying (e.g., a different version of \texttt{sentence-transformers}), it may lead to retrieval mismatches or errors. Always ensure that the same model is used both during indexing and querying. \\
    
    \textbf{Model version issues} \ Using an incorrect or unsupported model version (e.g., referencing a non-existent version like \texttt{Llama 4.5}) can lead to failures during the model loading process. Always verify that the model version you are using is supported and available. \\
    
    \textbf{Overly general prompts} \ Prompts that are too broad or generic can result in vague or irrelevant responses from the model. Craft precise and targeted prompts to ensure more accurate and relevant answers. \\
    
    \textbf{Ignoring context} \ Language models can generate incorrect or hallucinated responses when the retrieved documents lack the necessary information, leading the model to fill in gaps inaccurately. Always ensure sufficient context is provided in the query. \\
    
    \textbf{Memory leaks} \ Extended use of FAISS and OLlama in continuous loops without proper memory management can result in memory leaks, gradually consuming system resources. Monitor memory usage and free up resources after each loop to avoid system slowdowns. \\
    
    \textbf{Model re-initialization} \ Reloading or re-initializing models unnecessarily can slow down your system. Reuse initialized models whenever possible to improve system efficiency and reduce overhead. \\
\end{tabularx}
\end{tcolorbox}
\vspace{1em}
OLlama (Llama 3.1) is a local language model that runs entirely on the user's machine, ensuring data privacy and faster response times depending on the system's hardware. The accuracy of its outputs depends on the quality of its training data, and it can be further improved by fine-tuning with domain-specific knowledge. Fine-tuning involves retraining the model with specialized datasets, allowing it to internalize specific organizational knowledge for more precise and relevant responses. This process keeps the model updated and tailored to the user's needs while maintaining privacy.\\

\end{enumerate}

%%% Local Variables:
%%% mode: latex
%%% TeX-master: "../main.tex"
%%% End:

%% file: sections/05_Evaluation.tex
\section{Preliminary Evaluation of the Guide}

\subsection{Feedback Process Overview}

This experience report underwent an informal evaluation process aimed at gathering feedback for the section: \textbf{Using OpenAI's Assistant API : GPT Series}.[{\ref{OpenAI's Assistant API}}] Although the feedback session was not formally structured, it still provided valuable insights that helped validate the ideas presented in this section and refine the guide based on it. The feedback gathered from participants demonstrates that the workshop was successful. A majority of attendees were able to follow the provided guide and successfully implemented their RAG models by the end of the session.

\subsection{Participants}

\begin{figure}[ht]
    \centering
    % First image on the left
    \begin{subfigure}{0.45\textwidth}
        \centering
        \includegraphics[width=\linewidth]{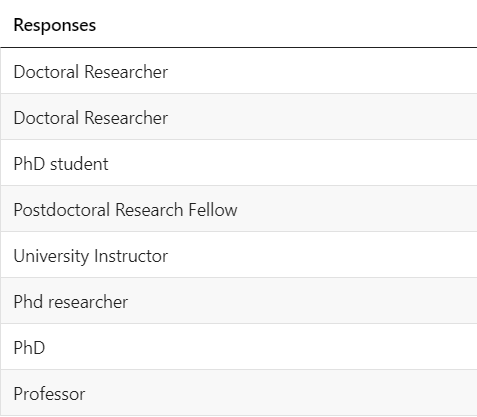}
        \caption{Job Titles of Participants}
        \label{fig:expertise}
    \end{subfigure}
    \hfill
    % Second image on the right
    \begin{subfigure}{0.45\textwidth}
        \centering
        \includegraphics[width=\linewidth]{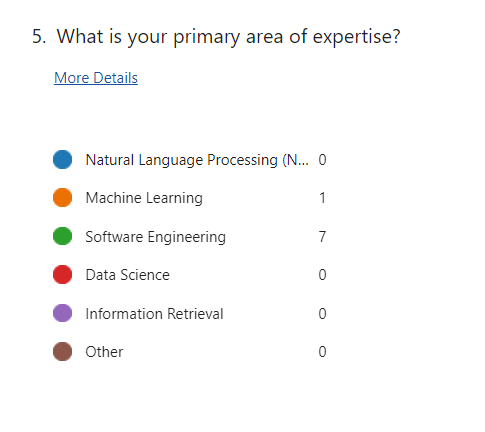}
        \caption{Primary Area of Expertise}
        \label{fig:job_titles}
    \end{subfigure}
    \caption{Demographic Information from Participants}
    \label{fig:demographics}
\end{figure}
\noindent
We collected feedback from a small but diverse group of participants during a workshop. A total of 8 individuals completed a demographics form, which provided us with an understanding of the participants’ backgrounds and technical expertise. The group consisted of individuals with varying levels of experience in machine learning, natural language processing (NLP), and using tools for Retrieval Augmented Generation (RAG). The participants although had familiarity with Python language and OpenAI models.

\subsection{Key Feedback Points}

During the session, participants shared their thoughts on how much their understanding of RAG systems improve after the workshop, which aspect of the workshop did they find most valuable, challenges they faced and what suggestions or comments they want to provide for future improvement.Below are the key points highlighted by the participants.

\begin{figure}[h!]
  \centering
  \includegraphics[width=1\textwidth]{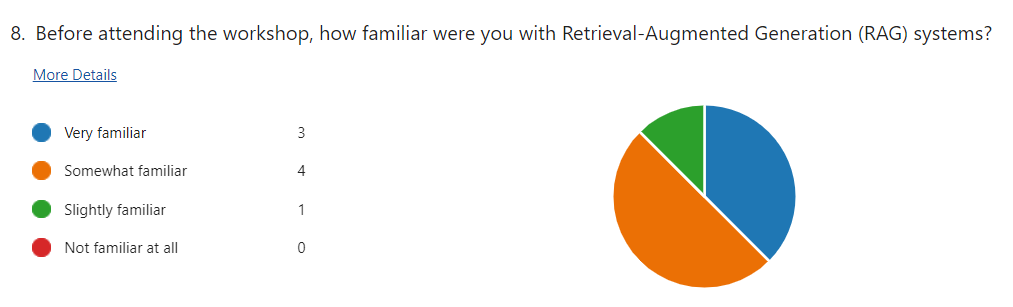}
  \caption{Participants' Familiarity with RAG Systems.}
  \label{fig:DemoPrior}
\end{figure}

Prior to attending the workshop, the majority of participants reported a reasonable level of familiarity with RAG systems.This indicated that the audience had a foundational understanding of the concepts presented, allowing for more in depth discussions during the workshop. After the workshop, there was a notable improvement in participants' understanding of RAG systems.

\begin{figure}[h!]
  \centering
  \includegraphics[width=1\textwidth]{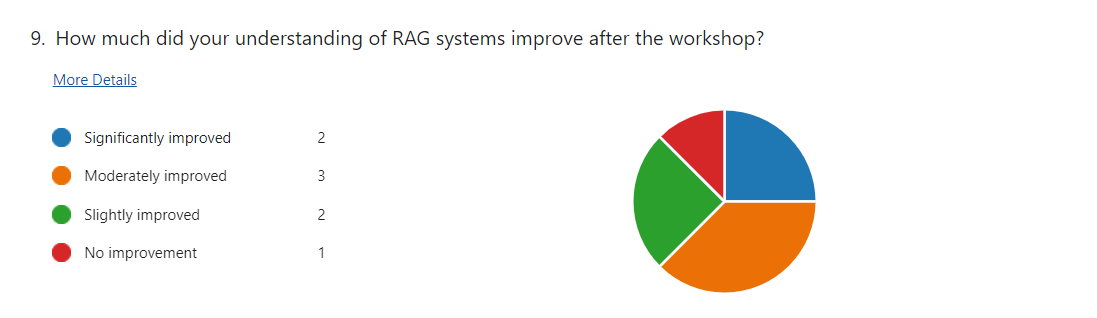}
  \caption{Participants' Improvement in Understanding RAG Systems.}
  \label{fig:DemoAfter}
\end{figure}

\begin{figure}[h!]
  \centering
  \includegraphics[width=1\textwidth]{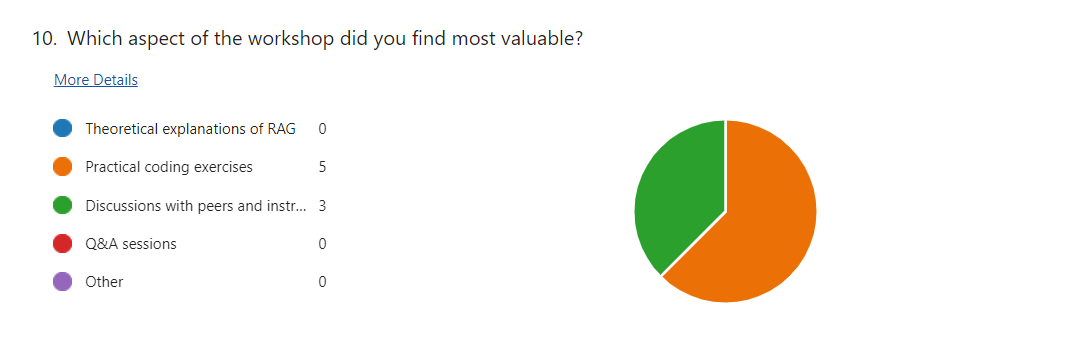}
  \caption{Most Valuable Aspects of the Workshop.}
  \label{fig:Demo}
\end{figure}

The majority of participants highlighted the practical coding exercises as the most valuable aspect of the workshop, which helped them better understand the implementation of RAG systems. Additionally, several participants mentioned the discussions with peers and instructors as a key takeaway. 

\subsection{Incorporating Feedback to Improve the Guide}

The evaluation also revealed opportunities for improvement of guide, particularly in enhancing the clarity of instructions and streamlining the implementation process. The most common issues raised were technical, most of them related to copying from PDF file generating errors as number of lines. In addition to that, we implemented error handling to throw meaningful errors to the user in the code snippets provided to seamlessly run the code.  

\begin{figure}[h!]
    \centering
    \includegraphics[width=\textwidth]{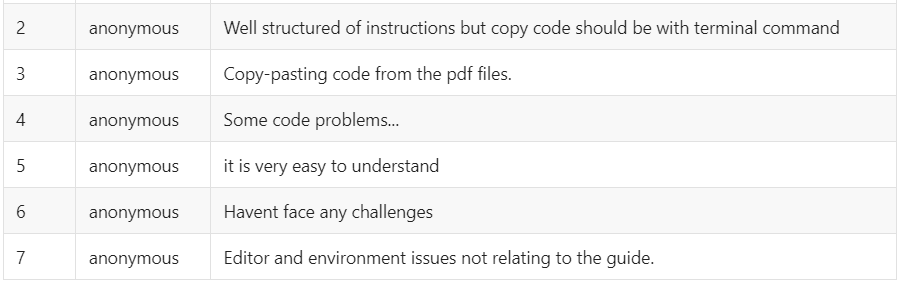}
    \caption{Feedback on challenges faced during the implementation of the guide}
    \label{fig:challenges_feedback}
\end{figure}

\begin{figure}[h!]
    \centering
    \includegraphics[width=\textwidth]{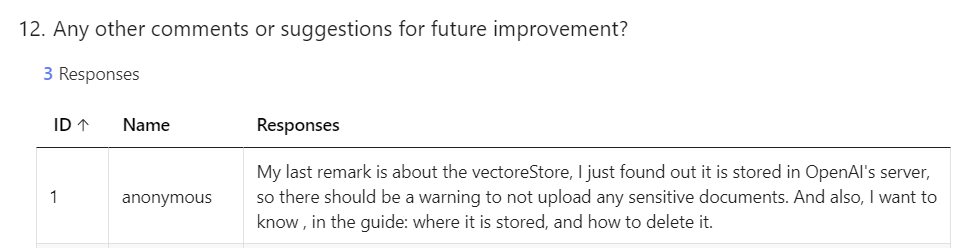}
    \caption{Comments and suggestions for improving the guide}
    \label{fig:improvement_suggestions}
\end{figure}

Several participants also provided suggestions for future improvement. One notable suggestion was to include a warning regarding sensitive data in OpenAI vector store. The detailed comment are shown in Fig[{\ref{fig:improvement_suggestions}}]. 

\vspace{1em}
In conclusion, the evaluation process proved valuable in validating the approach outlined in the guide. By testing it in a hands on workshop environment, and in an open discussion sessions of how the developed RAG models improved trustworthiness in specific scenarios, we were able to collect meaningful feedback and directly address areas of difficulty faced by practitioners. The feedback driven improvements have not only made the guide more user friendly but also demonstrated the importance of continuous iteration based on real world use.

%% file: sections/06_Discussion.tex
\section{Discussion}

Practitioners in fields like healthcare, legal analysis, and customer support, often struggle with static models that rely on outdated or limited knowledge. RAG models provide practical solutions with pulling in real time data from provided sources. The ability to explain and trace how RAG models reach their answers also builds trust where accountability and decision making based on real evidence is important.

In this paper, we developed a RAG guide that we tested in a workshop setting, where participants set up and deployed RAG systems following the approaches mentioned. This contribution is practical, as it helps practitioners implement RAG models to address real world challenges with dynamic data and improved accuracy. The guide provides users clear, actionable steps to integrate RAG into their workflows, contributing to the growing toolkit of AI driven solutions.

With that, RAG also opens new research avenues that can shape the future of AI and NLP technologies. As these models and tools improve, there are many potential areas for growth, such as finding better ways to search for information, adapting to new data automatically, and handling more than just text (like images or audio). Recent advancements in tools and technologies have further accelerated the development and deployment of RAG models. As RAG models continue to evolve, several emerging trends are shaping the future of this field.

\begin{enumerate}
    \item \textbf{Haystack}: An open-source framework that integrates dense and sparse retrieval methods with large-scale language models. Haystack supports real-time search applications and can be used to develop RAG models that perform tasks such as document retrieval, question answering, and summarization \cite{haystack2023}.
    
    \item \textbf{Elasticsearch with Vector Search}: Enhanced support for dense vector search capabilities, allowing RAG models to perform more sophisticated retrieval tasks. Elasticsearch’s integration with frameworks like Faiss enables hybrid retrieval systems that combine the strengths of both dense and sparse search methods, optimizing retrieval speed and accuracy for large datasets\cite{elasticsearch2023}.

    \item \textbf{Integration with Knowledge Graphs}: Researchers are exploring ways to integrate RAG models with structured knowledge bases such as knowledge graphs. This integration aims to improve the factual accuracy and reasoning capabilities of the models, making them more reliable for knowledge-intensive tasks\cite{xiong2024knowledge}.
    
    \item \textbf{Adaptive Learning and Continual Fine-Tuning}: There is a growing interest in adaptive learning techniques that allow RAG models to continuously fine-tune themselves based on new data and user feedback. This approach aims to keep models up-to-date and relevant in rapidly changing information environments\cite{liang2023best}.
    
    \item \textbf{Cross-Lingual and Multimodal Capabilities}: Future RAG models are expected to expand their capabilities across different languages and modalities. Incorporating cross-lingual retrieval and multimodal data processing can make RAG models more versatile and applicable to a wider range of global and multimedia tasks\cite{chowdhury2024cross}.
\end{enumerate}

Future research will likely focus on enhancing their adaptability, cross-lingual capabilities, and integration with diverse data sources to address increasingly complex information needs.

%%% Local Variables:
%%% mode: latex
%%% TeX-master: "../main.tex"
%%% End:

%% file: sections/07_Conclusions.tex
\section{Conclusions}
The development of Retrieval Augmented Generation (RAG) systems offers a new way to improve large language models by grounding their outputs in real-time, relevant information. This paper covers the main steps for building RAG systems that use PDF documents as the data source. With clear examples and code snippets, it connects theory with practice and highlights challenges like handling complex PDFs and extracting useful text. It also looks at the options available, with examples of using proprietary APIs like OpenAI’s GPT and, as an alternative, open-source models like Llama 3.1, helping developers choose the best tools for their needs.

By following the recommendations in this guide, developers can avoid common mistakes and ensure their RAG systems retrieve relevant information and generate accurate, fact-based responses. As technology advances in adaptive learning, multi-modal capabilities, and retrieval methods, RAG systems will play a key role in industries like healthcare, legal research, and technical documentation. This guide offers a solid foundation for optimizing RAG systems and extending the potential of generative AI in practical applications.

%%% Local Variables:
%%% mode: latex
%%% TeX-master: "../main.tex"
%%% End: